\documentclass[journal, 2p]{elsarticle}

\usepackage{lipsum}

\makeatletter
\def\ps@pprintTitle{%
 \let\@oddhead\@empty
 \let\@evenhead\@empty
 \def\@oddfoot{\rightline{\today}}%
 \let\@evenfoot\@oddfoot}
\makeatother

\usepackage[utf8]{inputenc}
\usepackage[T1]{fontenc}
\usepackage{lmodern} 
\usepackage{appendix}

\usepackage{siunitx}

\usepackage{algorithm}
\usepackage{algpseudocode}
\usepackage{comment}
\newcommand{\dA}{\mathbf{\Delta}} %\mathbf{\Delta}}

\newcommand{\sign}{\mbox{sign}}
\newcommand{\argmin}{\mbox{argmin}}
\newcommand{\argmax}{\mbox{argmax}}
\newcommand{\R}{\mathbb{R}}

\newcommand{\ith}{ ^{\text{th}} }

\newcommand{\oner}{\mathbf{1}}
\newcommand{\tand}{\text{ and }}
\newcommand{\tor}{\text{ or }}
\newcommand{\mbyn}{{m \times n}}

\newcommand{\bfeta}{\boldsymbol{\eta}}
\newcommand{\by}{\mathbf{y}}
\newcommand{\bx}{\mathbf{x}}
\newcommand{\bg}{\mathbf{g}}
\newcommand{\bh}{\mathbf{h}}
\newcommand{\bt}{\mathbf{t}}

\newcommand{\bq}{\mathbf{q}}
\newcommand{\bA}{\mathbf{A}}
\newcommand{\bC}{\mathbf{C}}
\newcommand{\bG}{\mathbf{G}}
\newcommand{\bH}{\mathbf{H}}
\newcommand{\bS}{\mathbf{S}}
\newcommand{\bI}{\mathbf{I}}
\newcommand{\bM}{\mathbf{M}}
\newcommand{\bD}{\mathbf{D}}
\newcommand{\bLam}{\boldsymbol{\Lambda}}

\newcommand{\xtrue}{\mathbf{x}_{\text{TRU}}}
\newcommand{\xmle}{{\mathbf{x}}_{\text{AML}} }
\newcommand{\xols}{{\mathbf{x}}_{\text{OLS}}}
\newcommand{\xtls}{{\mathbf{x}}_{\text{TLS}}}
\newcommand{\xest}{{\mathbf{x}}_{\text{EST}}}
\newcommand{\diag}{\text{diag}}
\newcommand{\bzero}{\mathbf{0}}
\newcommand{\dtx}{\delta  \bt \bx^T}

\newcommand{\p}{\mathbf{\mathcal{P}}}
\newcommand{\csch}{\text{csch}}

\usepackage{url}

\usepackage{graphicx}
\usepackage{amsmath,amsfonts,amssymb, amsthm}
\usepackage[usenames,dvipsnames,svgnames]{xcolor} % use like \textcolor{red}{..} or {\color{red} ...}
\usepackage{booktabs}
\usepackage{siunitx} % see https://tex.stackexchange.com/a/458965 and http://ctan.math.washington.edu/tex-archive/macros/latex/contrib/siunitx/siunitx.pdf
\usepackage{etoolbox}

\begin{document}

\begin{frontmatter}
 \title{Approximate maximum likelihood estimators for linear regression with design matrix uncertainty}

\cortext[cor1]{Corresponding author}
\author{Richard J. Clancy\corref{cor1}}
\ead{richard.clancy@colorado.edu}

\author{Stephen Becker}
\ead{stephen.becker@colorado.edu}

\address{Department of Applied Mathematics at the University of Colorado, Boulder, CO 80309}

%\vspace*{-1in}

\begin{abstract}
	In this paper we consider regression problems subject to arbitrary noise in the operator or design matrix. This characterization appropriately models many physical phenomena with uncertainty in the regressors. Although the problem has been studied extensively for ordinary/total least squares, and via models that implicitly or explicitly assume Gaussianity, less attention has been paid to improving estimation for regression problems under general uncertainty in the design matrix. To address difficulties encountered when dealing with distributions of sums of random variables, we rely on the saddle point method to estimate densities and form an approximate log-likelihood to maximize. We show that the proposed method performs favorably against other classical methods.
\end{abstract}

\begin{keyword}
	\noindent Least squares approximation \sep Quantization error \sep Maximum likelihood estimation \sep Saddle point approximation \sep Total least squares \sep Design matrix uncertainty \sep Operator uncertainty 
\end{keyword}

\end{frontmatter}
% --------------------------------------

%%%%%%%%%%%%%%%%%%%%%%%%%%%%%%%%%%%%%%%%%%%%%%%%%%%%%%%%%%%%%%%%%%%%%
%%%%%%%%%%%%%%%%%%%%%%%% INTRODUCTION %%%%%%%%%%%%%%%%%%%%%%%%
\section{Introduction} \label{sec:intro}
We consider the linear regression problem subject to uncertainty in the operator or data/design matrix given by the generative model 
\begin{equation} \label{eq:generative_model}
    \by = \bG \bx + \bfeta,
\end{equation}
where $\bG \in \R^\mbyn$ and $\bfeta \in \R^m$ are random variables with independent components, and $\bx \in \R^n$ is a fixed but unknown parameter vector. Typical causes of operator uncertainty are sampling error, measurement error, human error, modeling error, or rounding error. Our focus is on the over-determined case when $m > n$. We assume distributional knowledge of both $\bG$ and $\bfeta$. Our goal is to recover an estimate for $\bx$ given observations of $\by$ using a maximum likelihood estimation (MLE) framework. 

We have several motivating examples in mind. The first involves quantization error where design matrix elements are rounded to a fixed decimal place. Such instances naturally arise during digitization and can be modeled as Berkson error. Another typical case is for surveys or ratings where respondents answer questions on a Likert scale to estimate a response variable, e.g., suitability of an applicant for a particular job. A continuum of preferences are forced to take integer values. Furthermore, survey responses are subjective in nature introducing more uncertainty. 

A second example involves effectively estimating $\bx$ when the design matrix is subject to floating point error. This might be encountered when data is roughly transcribed  or the number of significant figures (digits in the mantissa) stored on a drive are limited for memory savings. Floating point error is a generalization of rounding error. 

The third example is for clipping, i.e., we observe $\bH =  \sign(\bG) \cdot \max(\bG, \gamma)$ where $\max(\cdot, \gamma)$ operates element-wise and $\gamma$ is a clipping threshold. If $\bG$ is drawn from a double exponential, then the uncertainty in elements of $\bH$ taking extreme values of $\pm \gamma$ will be distributed exponentially (up to a sign). Given the heavy tails of exponentials, it is desirable to incorporate operator uncertainty in the problem formulation.   

Our decision to focus on MLEs is supported by the fact that well known regression formulations are in actuality MLEs for additive noise, that is, for $\by = \bA \bx +  \bfeta$ where $\bfeta$ is random but $\bA$ is known. In particular, $\argmin_x \|\bA \bx - \by\|_2^2$ (ordinary least squares), $\ \argmin_x \|\bA \bx - \by\|_1$ (least deviation regression), and $\argmin_x \|\bA \bx - \by\|_{\infty}$ (minimax regression) correspond to the MLEs for Gaussian, double exponential, and uniform noise, respectively.

These MLE regression problems rely on knowledge of the vector $\by$'s joint probability density function (PDF). Note that each component of $\by$ in \eqref{eq:generative_model} is the sum of scaled random variables, i.e., $y_i = \bg_i^T \bx + \eta_i = \sum_{i=1}^n G_{ij}x_j + \eta_i$ where $\bg_i^T$ is the $i\ith$ row of $\bG$ and subscripts denote the component of the corresponding vector. Despite the innocuous form, sums of random variables are difficult to work with: individual PDFs must be convolved to obtain a PDF for their sum. 

One option is to ignore uncertainty in $\bG$ altogether, focusing on additive noise in $\by$ alone as done in ordinary least squares (OLS). Although reasonable when uncertainty in the design matrix is small, this simplification often fails in practice. Total least squares (TLS) was developed to resolve the asymmetry in uncertainty between the design matrix and the measurement vector. It is often used with the errors-in-variables model (EIV). TLS solves the problem 
\begin{equation} \label{eq:TLS}
    \min_{\bx,\bfeta, \dA}  \quad \big\| [\dA, \  \ \bfeta] \big\|_F \quad \text{subject to} \quad  (\bA + \dA)\bx = \by + \bfeta 
\end{equation}
where $[\dA, \bfeta] \in \R^{m \times (n+1)}$ is minimized with respect to the Frobenius norm. The model supposes that for an observed $\bA$, there is true deterministic $\bA_{\text{TRUE}} = \bA + \dA^*$ where $\dA^*$ solves problem \eqref{eq:TLS}. Golub and Van Loan showed that TLS can be solved via a singular value decomposition with a closed form solution $\hat \bx_{\text{TLS}} = (\bA^T\bA - \sigma_{n+1}^2 \bI)^{-1} \bA^T \by$ \cite{golub1980tls} with $\sigma_{n+1}$ being the smallest singular value of $[\bA, \by]$. 
The TLS solution coincides with the MLE for a deterministic $\bA_{\text{TRUE}}$ with independent, identically, distributed additive Gaussian noise \cite{van1991total, fuller1987measurement}. 
An analytic solution is appealing but it suffers from a deterioration in conditioning. 

As an alternative to TLS, Wiesel, Eldar, and Yeredor \cite{wiesel2008linear} devised a MLE when the design matrix is a random variable with all uncertainty normally distributed. We use the same generative model in our setup, but allow for noise from general distributions. By exploiting properties of Gaussian distributions, they formed a likelihood function (LF), showed its equivalence to a (de)regularized least squares problem, and provided algorithms to find the estimator. Gaussiantity is central to their analysis, simplifying otherwise intractable calculations. 

Efforts have been made to move away from Gaussian noise through robust optimization where the goal is to generate estimates impervious to perturbations in the observed data. Many robust optimization problems are cast in a minimax form \cite{CaramanisLasso, goldfarb2003robust, el1997robust, BertsimasCaramanisRO, becker2020robust}. Typically, an uncertainty set $\mathcal{U}$ and objective function $f$ are specified, then the aim is to solve $\min_x  \left\{ \max_{U \in \mathcal{U}} f(x,U) \right\}$. There are a number of drawbacks to the robust framework; in particular, estimates tend to be overly conservative. Furthermore, these methods discard distributional knowledge of the noise focusing instead on set geometry. 

To make progress on the MLE in general, we require a method to efficiently construct a probability density. Although this is a difficult task for all but a few special distributions, there is hope. Given certain regularity conditions, work in the Laplace domain is possible through a bilateral transformation. Rather than working with PDFs directly, we can use their moment generating functions (MGF). The ideas in this paper rely on two important properties of MGFs: 
\begin{enumerate}
    \item the MGF for a sum of independent random variables is the product of their individual MGFs, i.e., $M_{X+Y}(t) = M_X(t)M_Y(t)$ and
    \item the MGF uniquely characterizes a random variable as its PDF does.
\end{enumerate}
Not all random variables permit an MGF (e.g., the Cauchy distribution lacks one) but many distributions of practical interest do, including Bernoulli, binomial, Poisson, uniform, Gaussian, exponential distributions, as well as all distributions with bounded support.
It is assumed throughout this paper that all random variables discussed have MGFs.

Using moment generating functions, we can easily specify the distribution for linear combinations of random variables found in the regression problem. Recovery of a PDF for use in a LF is possible, in theory, through inversion of the MGF but is often difficult in practice. Instead, by using the MGF, we can employ the saddle point method to estimate the PDF and form an approximate LF. We then maximize the approximate LF to recover an estimate of $\bx$ in Eq. \ref{eq:generative_model} accounting for uncertainty in matrix $\bG$. 

The saddle point method is a generalization of Laplace's method and was first used by Debye to study Bessel functions of high order \cite{debye1909naherungsformeln} then by Watson for statistical mechanics \cite{watson1995treatise}. It was extended by Daniels in his seminal paper \cite{daniels1954saddlepoint} to estimate the PDF of sample means. Barndorff-Nielsen and Cox \cite{barndorff1979edgeworth} and Luggannani and Rice \cite{lugannani1980saddle} generalized his work for independent sums of random variables. Spady \cite{spady1991saddlepoint} discussed saddle point approximations for linear regression problems but focused on the distribution of a user specified ``estimating equation'' such as $\nabla_{\bx} \|\bG \bx - \by \|^2$ or the subdifferential of $\|\bG \bx - \by\|_1$. Stawdermann, Casella, and Wells \cite{strawderman1996practical} used the saddle point method to approximate distributions of MLEs in the regression problem, but focused on its statistical properties. A number of works use Laplace's approximation (the real analog to the saddle point method) for parameter estimation in general linear models but most applications are concerned with longitudinal scientific studies and assume Gaussianity throughout \cite{vonesh1996note, ko2000correcting, battauz2011laplace}. 

In this paper, we derive an approximate MLE for point estimation in linear regression problems with uncertainty in both the measurement vector and design matrix. The approximate MLE is based on easily computed univariate MGFs and can accommodate noise from general distributions. We also provide an expression for the corresponding gradient allowing for the use of off-the-shelf first-order optimization solvers and discuss algorithmic considerations. To our knowledge, the saddle point method has not been used in this setting before.

In Section \ref{sec:background}, we provide the mathematical background for the proposed method. We formulate and present our approximate (log) likelihood function in Section \ref{sec:likelihood_function} then introduce algorithms to solve the approximate MLE problem in Section \ref{sec:algorithms}. We motivate its utility through illustrative examples in Section \ref{sec:example} then present results of numerical experiments in Section \ref{sec:experiments}.

%%%%%%%%%%%%%%%%%%%%%%%%%%%%%%%%%%%%%%%%%%%%%%%%%%%%%%%%%%%%%%%%%%%%%
%%%%%%%%%%%%%%%%%%%%%%%% BACKGROUND %%%%%%%%%%%%%%%%%%%%%%%%
\section{Background} \label{sec:background}
Maximum likelihood estimation is one of the most commonly used techniques in statistics. MLEs require
the construction of a LF which varies in difficulty. In the regression problem, we are concerned with sums of random variables and their corresponding densities. Although there are several well known distributions for sums of random variables, such as sums of normals, $\chi^2$'s, exponentials, etc., the majority of distributions do not enjoy elegant forms. Furthermore, many densities for sums require them to be identical, severely curtailing their usefulness; linear combinations are out of reach. Density construction is possible through convolution but presents serious difficulties in practice. 

Rather than manipulating densities directly, we can work with their moment generating functions (MGF). The idea is closely related to Fourier analysis where a convolution in time becomes multiplication in the frequency domain. The MGF
is simply a bilateral Laplace transform of the PDF, $f_X$, given by 
\begin{equation} \label{eq:mgf}
    M_X(t) = \mathbb{E}\left(  e^{tX} \right) = \int_{-\infty}^{\infty} e^{tx}f_X(x) \, dx.
\end{equation}
The existence of an MGF is not guaranteed, but when it does exist, it uniquely characterizes the random variable. We exploit this idea by encoding the random variable's statistics in the moment generating function, then use it to construct an approximate LF. 

To illustrate, let $U \sim \text{Uniform}(0,1)$ and $Z \sim \mathcal{N}(0, 1)$ with known densities of $f_U(u) = I_{(0,1)}(u)$ and $f_Z(z) = (1/\sqrt{2 \pi}) \exp{\{-z^2/2\}}$, respectively (we use $I_\mathcal{A}$ to denote indicator function over $\mathcal{A}$). The density of $U+V$ is 
\[
f_{U+Z}(y) = \frac{1}{\sqrt{2\pi}}\int_0^1 e^{-\frac{(y-s)^2}{2}} \, ds
\]
which has no analytic form. In contrast, the MGFs are $M_U(t) = (\exp{\{t\}} - 1)/t$ and $M_Z(t) = \exp{\{-t^2/2\}}$. The MGF of their sum is 
\[
M_{U+Z}(t) = \frac{(e^t - 1)e^{-t^2/2}}{t}.
\] 
The MGF appears more complicated but is \textit{exact}. In contrast, the PDF requires evaluation of an integral at each point and relies on approximation through numerical integration. Use of the PDF presents no real difficulty in our example, but as the number of random variables in the sum increase or distributions vary, construction and/or evaluation of the PDF, as well as finding the gradient of its associated likelihood, becomes problematic.

Using the generative model, we aim to form a likelihood function, $L$, based on observations of $\by$. We assume that both $\bG$ and $\bfeta$ are component-wise independent and that their distributions are known. We cannot observe $\bG \tor \bfeta$ directly, but instead observe $\by$ which is a function of both. For notational simplicity, let $\bG \sim \p_\bG \tand \bfeta \sim \p_{\bfeta}$ with $\p$ denoting the respective distributions. The likelihood will be a function of $\bx$ and depend on $\p_{\bG}, \p_{\bfeta}, \tand \by$ given as
\begin{equation} \label{eq:jointdensity}
    L(\bx) = p(\by; \, \bx, \p_{\bG}, \p_{\bfeta}). 
\end{equation} 
We require a closed form expression for $p$, the PDF of $\by = \bG \bx + \bfeta$, which is a component-wise weighted sum of random variables. As discussed earlier, it is difficult to compute in general. 

Given the trouble of forming an exact PDF, we focus on approximation methods. One option is to randomly sample the distribution then form a kernel approximation \cite{rosenblatt1956, parzen1962estimation}. Unfortunately, kernel density estimation requires many samples to adequately capture the structure of the distribution and its accuracy is influenced by the user's choice of a kernel, making in unsuitable for our regression problem. Another popular choice is to use an Edgeworth series expansion \cite{hall2013bootstrap} where a polynomial approximation for the density is used that matches the first several cumulants of the true density. A major drawback of the method is that it introduces false critical points unrelated to the actual density. Since the ultimate objective is to maximize the LF, the addition of phantom critical points creates otherwise avoidable difficulties. Furthermore, the Edgeworth expansion can take negative values which violates the properties of a PDF. At this point, we turn our attention to the saddle point method to approximate our PDF based on the exact MGF.

\begin{figure}
    \centering
    \includegraphics[width=.6\textwidth]{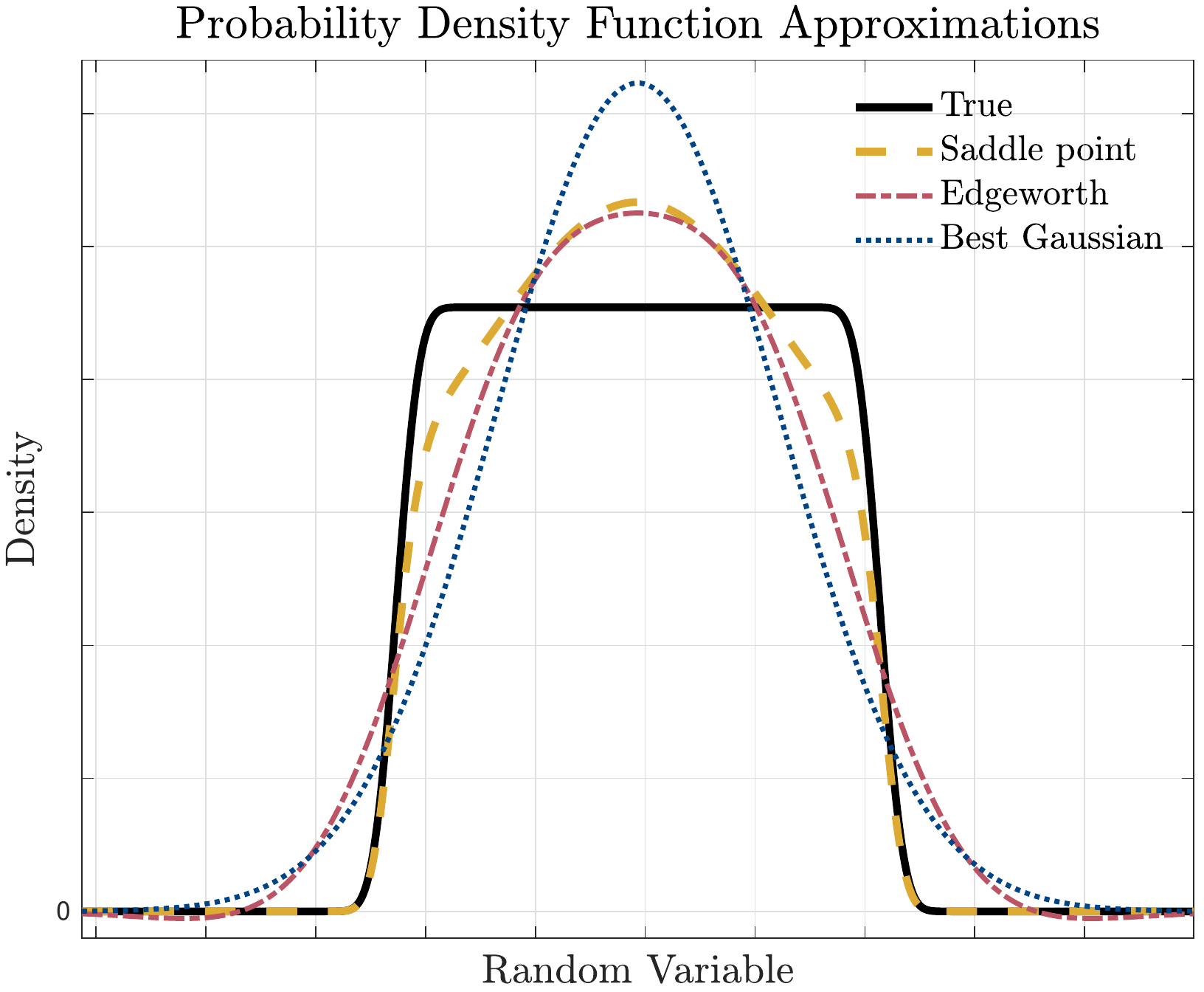}
    \caption{Example of the true density for $y$ and several approximations when $y = \bg^T \bx + \eta$ with $\bg$ uniform, $\bx$ from a Cauchy distribution, and $\eta$ normal. Note that $\bx$ is drawn randomly to start but fixed for pdf of $y$. The saddle point approximation fits distributional tails exceptionally well. Note that Edgeworth expansion takes negative values and introduces phantom critical points. The best Gaussian has a matching mean and variance as the true distribution.}
    \label{fig:densities}
\end{figure}

%%%%%%%%%%%%%%%%%%%%%%%% SADDLE POINT APPROXIMATION %%%%%%%%%%%%%%%%%%%%%%%%
\subsection*{Saddle point approximation} 
We outline the principle behind the saddle point method here with an informal treatment. The method is closely related to the method of steepest descent and the stationary-phase method.
For rigorous arguments, the interested reader may refer to the original paper \cite{daniels1954saddlepoint} or one of the excellent overview articles \cite{reid1988saddlepoint, goutis1999explaining, huzurbazar1999practical}. Our aim is to give a general idea of the method for illustrative purposes. 

We start with a well known theorem in probability stating that for a random variable $A$, if the MGF $M_A(iw)$ is integrable (with $i = \sqrt{-1}$),  then it has a PDF given by 
\[
    f_A(\alpha) = \frac{1}{2\pi}\int_{-\infty}^{\infty} M_A(iw) \, e^{-iw\alpha} \, dw.
\]
This is the inverse Fourier transform of the MGF with a complex argument. By a change of variable, $t = iw$, the integral becomes
\[
    f_A(\alpha) = \frac{1}{2 \pi i}\int_{-i\infty}^{i \infty} M_A(t) \, e^{-t\alpha} \, dt = \frac{1}{2 \pi i} \int_{-i\infty}^{i \infty}  e^{\ln M_A(t) -t\alpha} \, dt.
\]
For notational simplicity, the \textit{cumulant generating function} (CGF) is defined as $K_A(\alpha) = \ln M_A(\alpha)$. Assuming reasonable regularity conditions and by the Closed Curve Theorem, we can rewrite an equivalent integral with our contour translated by $\tau \in \R$ (to be fixed shortly) along the real axis such that 
\[
    f_A(\alpha) = \frac{1}{2 \pi i}\int_{\tau-i\infty}^{\tau + i \infty} e^{K_A(t)-t\alpha} \, dt. 
\]
To approximate the integral, the exponent is Taylor expanded about its maximum value on the contour. This agrees with intuition; the approximation will be accurate where the integral's mass lies. Points where the expansion is less accurate are effectively down-weighted through exponentiation. The quality of the approximation is contingent on the nature of $K_A(t) - t \alpha$, but is favorable for sums of random variables. 

The integrand takes its maximum value along the contour at a critical point, that is, when $K_A'(t) - \alpha = 0$. It so happens that when $\alpha$ is in the support of the PDF for $A$, there is a unique real root for the preceding equation \cite{daniels1954saddlepoint}. We call this point $t_0 \in \R$. Since the maximum occurs on a contour parallel to the imaginary axis, a minima occurs at the same point along the real axis because the exponent is analytic and must fulfill the Cauchy-Riemann equations, hence $K''_A(t_0) > 0$. Taylor expanding about $t_0$ and shifting the contour to pass through it, i.e., $\tau = t_0$, gives, 
\begin{align*}
    f_A(\alpha) & \approx \frac{1}{2 \pi i}\int_{t_0-i\infty}^{t_0+ i \infty} \exp \bigg\{(K_A(t_0)-t_0\alpha) \\
    & \qquad \qquad \qquad \quad \ + (K_A'(t_0)-\alpha)(t-t_0) + \frac{1}{2}K_A''(t_0) (t-t_0)^2   \bigg\} \, dt \\
    & \quad \\
    & =  \ \frac{1}{2 \pi} \exp\{K_A(t_0) - t_0 \alpha\} \int_{t_0-i\infty}^{t_0 + i \infty} \frac{1}{i} \exp \left\{\frac{1}{2}K_A''(t_0) (t-t_0)^2 \right\} \, dt 
\end{align*} 
Through successive substitutions, the integrand can be rewritten as a Gaussian function and integrated over $\R$ yielding an approximation of the PDF for $A$,
\begin{equation} \label{eq:saddlepointapprox}
    f_A(\alpha) \approx \tilde f_A(\alpha) = \sqrt{\frac{1}{2\pi K_A''(t_0)}}  \ e^{K_A(t_0) - t_0 \alpha}
\end{equation}
where $t_0$ depends on $\alpha$ through the equation $K'_A(t_0) - \alpha = 0$. The function $\tilde f_A(\alpha)$ is known as the \textit{saddle point approximation} for $f_A$ at point $\alpha$. Higher order approximations can be achieved by including more terms in the Taylor expansion, but we limit our discussion to quadratics.   

%%%%%%%%%%%%%%%%%%%%%%%% NOTATION %%%%%%%%%%%%%%%%%%%%%%%%
\subsection*{Notation}
Bold lower/upper case letters denote vectors/matrices, while unbolded lower-case letters are scalars. For a matrix $\bG$, the $i\ith$ row and $j\ith$ column are given by $\bg_i^T$ and $\bg_j$, respectively. A vector $\by$ is composed of components $y_i$ with the subscript denoting the index of an element. We write the joint CGF of a random variable $\by \in \R^m$ as $K_{\by}(\bt) = [K_{y_1}(t_1), \, ..., \, K_{y_m}(t_m)]^T$ and its derivatives $K_{\by}^{(p)} (\bt) = [ \frac{\partial^p}{\partial t_1^p} K_{y_1}(t_1), \, ..., \, \frac{\partial^p}{\partial t_m^p} K_{y_m}(t_m)]^T$. Unless otherwise noted, the variables $\bt$ and $t_i$ are reserved to denote solutions to equation $K'_{\bG \bx + \bfeta}(\bt) - \by=\bzero$ where $t_i$ is the solution to the $i\ith$ component of the vector equation.

A scalar function $f: \R \rightarrow \R$ applied to a matrix $\bA$ acts component-wise, i.e., $[f(\bA)]_{ij} = f(\bA_{ij})$. The Hadamard (or element-wise) product/quotient is denoted by $\odot / \oslash$. For a vector $\by \in \R^m$, the matrix $\diag(\by) \in \R^{m \times m}$ has the components of $\by$ along its diagonal and zeros elsewhere. A vector of ones and zeros is denoted by $\oner \tand \bzero$, respectively.

\begin{comment}
Using the derivation above, we can formulate an approximation for the joint density as seen in \eqref{eq:jointdensity} and \eqref{eq:prodjointdensity} given by 
\begin{equation} \label{eq:approxjointdensity}
    p(\by; \, \bx, \p_{\bG}, \p_{\bfeta}) \approx \prod_{i=1}^m \left(2\pi \, K_{\bg_i^T \bx + \eta_i}''(t_i)\right)^{-1/2}  \ \exp\left\{K_{\bg_i^T \bx + \eta_i}(t_i) - t_i \right \}
\end{equation}
Once again, $t_i$ satisfies the critical point equations $K'_{\bg_i^T \bx + \eta_i}(t_i) - y_i = 0$ for $ i \in \{1, \dots, \, m\}$.
\end{comment}

%%%%%%%%%%%%%%%%%%%%%%%%%%%%%%%%%%%%%%%%%%%%%%%%%%%%%%%%%%%%%%%%%%%%%
%%%%%%%%%%%%%%%%%%%%%%%% LIKELIHOOD FUNCTION %%%%%%%%%%%%%%%%%%%%%%%%
\section{Likelihood function} \label{sec:likelihood_function}
With an approximate density in hand, we can use \eqref{eq:jointdensity} and \eqref{eq:saddlepointapprox} to form a likelihood function (LF). Begin by noting that the joint density for a random vector with independent components can be rewritten as a product,
\begin{equation}
    p(\by; \, \bx, \p_{\bG}, \p_{\bfeta}) = \prod_{i=1}^m f_{Y_i}\left(y_i; \, \bx, \p_{\bG}, \p_{\bfeta} \right) = \prod_{i=1}^m f_{\bg_i^T \bx + \eta_i}(y_i),
\end{equation}
where $f_{Y_i}$ is the PDF for the $i\ith$ component. Since PDF construction is difficult for sums of random variables, we use the saddle point approximation and the LF from \eqref{eq:jointdensity} such that
\begin{equation}
    L(\bx) = p(\by; \, \bx, \p_{\bG}, \p_{\bfeta}) \approx \prod_{i=1}^m \tilde f_{\bg_i^T \bx + \eta_i} (y_i).
\end{equation}
We now define our \textit{approximate-likelihood function} (or approximate-LF) as 
\begin{equation} \label{eq:likelihood}
    \mathcal{L}(\bx) = \prod_{i=1}^m \left[ \left(2\pi \, K_{\bg_i^T \bx + \eta_i}''(t_i)\right)^{-1/2}  \ \exp\left\{K_{\bg_i^T \bx + \eta_i}(t_i) - t_i y_i\right\} \right]
\end{equation}
which assumes known measurements for $y_i$ and distributional knowledge of $\bG$ and $\bfeta$. Recall that $t_i$ solves $K'_{\bg_i^T \bx + \eta_i}(t_i) = y_i$.

As with any MLE problem, our goal is to find $\bx$ that maximizes the LF. For numerical stability, we focus on the log-likelihood function (log-LF), i.e., $\ell(\bx) = \ln \mathcal{L}(\bx)$, which has the same maximizer as \eqref{eq:likelihood} (we also omit the constant $2\pi$ since it doesn't impact the location of optima) and is given by 
\begin{equation} 
    \ell(\bx) = \sum_{i=1}^m  \left[ K_{\bg_i^T \bx + \eta_i}(t_i)  - \frac{1}{2} \ln\left(   K_{\bg_i^T \bx + \eta_i}''(t_i) \right) - t_i y_i \right].
\end{equation}

By exploiting the independence of components for both $\bG \tand \bfeta$ and using properties of MGFs, we can write $K_{\bg_i^T \bx + \eta_i}(t_i) = K_{\eta_i}(t_i) + \sum_{j=1}^n K_{G_{ij}}(x_j t_i)$. Since both $\eta_i$ and $G_{ij}$ have known CGFs, explicit calculation of $\ell(\bx)$ is easy and the log-LF is given by
\begin{align} \label{eq:elementwiseloglikelihood}
    \ell(\bx) = \sum_{i=1}^m  \left[ \rule{0cm}{.75cm} \right.  K_{\eta_i}(t_i) \ +\  &\left(\sum_{j=1}^n K_{G_{ij}}( t_i x_j) \right)  \nonumber \\
    \ -\ & t_i y_i - \frac{1}{2} \ln\left( K''_{\eta_i} (t_i) + \sum_{j=1}^n K''_{G_{ij}}(t_i x_j) \right)  \left. \rule{0cm}{.75cm} \right].
\end{align} 
We remind the reader that each $t_i$ is a function of $y_i \tand \bx$ through the critical point equation although we leave the dependence out for notational simplicity. If functional forms of $K_{\bG \bx + \bfeta}(\bt)$ and its derivatives are known, it is easier to work with the vectorized version, written as 
\begin{equation} \label{loglikelihood}
    \ell(\bx) = \oner^T \left( K_{\bG \bx + \bfeta}(\bt) - \frac 1 2 \ln \left( K''_{\bG \bx + \bfeta}(\bt)   \right) \right) - \bt^T \by.
\end{equation}
Functional forms for a particular problem are typically found by working directly with component-wise elements (Equation \ref{eq:elementwiseloglikelihood}) first.

Under the framework presented here, the observed matrix $\bG$ goes unused since all pertinent statistical information is embodied in the CGF. In practice, we envision the observed $\bG$ as the sample mean with noise distributed about it, as will be illustrated in Section \ref{sec:example}. If distributional parameters are unknown, such as variance for a Gaussian, we could treat them as variables in our log-LF. The actual formulation and gradient calculations would change only slightly, although the problem may have more spurious stationary points.

\subsection*{Log-likelihood gradient}
Although we have a log-LF, we must still maximize it. Most optimization algorithms rely on gradients so we focus on that computation now. To complicate matters, our log-LF depends on $\bt$ which is coupled to $\bx$ and, in general, has no closed form solution. To deal with this, we proceed with implicit differentiation as is done in the adjoint state method by treating $\bt$ as an independent variable. We recast our MLE problem as a mathematical program
\begin{align} \label{eq:optimization_problem}
    \underset{\bx, \bt}{\argmax} & \qquad \qquad \ell(\bx, \bt) \nonumber \\
    \text{subject to} & \quad K'_{\bG \bx + \bfeta}(\bt) - \by = \bzero.
\end{align} 
It follows from the chain rule that 
\begin{equation} \label{eq:dldx}
    \nabla_{\bx} \ell = \frac{\partial \ell}{\partial \bx} + \left( \frac{\partial \ell}{\partial \bt} \right)  \left( \frac{\text{d} \bt}{\text{d} \bx} \right).
\end{equation}
We follow the convention for Jacobians that the column indicates which variable is being differentiated and the row represents the component, i.e., $\left(\text{d} \bt / \text{d} \bx \right)_{ij} = \partial t_i  / \partial x_j$. By treating $\bx \tor \bt$ as fixed, it is straightforward to calculate the partial derivatives of $\ell(\bx, \bt)$. It remains to find $(\text{d} \bt / \text{d}\bx) \in \R^{m \times n}$.

To ease notation, let $\bq(\bx, \bt) = K'_{\bG \bx + \bfeta}(\bt) - \by$. Differentiating the constraint $\bq(\bx, \bt)=0$ gives
\begin{equation} \label{eq:dtdx}
    \frac{\partial \bq}{\partial \bx} + \left( \frac{\partial \bq}{\partial \bt} \right) \left( \frac{\text{d} \bt}{\text{d} \bx} \right) = \bzero \quad \iff \quad \frac{\partial \bt}{\partial \bx} = - \left( \frac{\partial \bq}{\partial \bt} \right)^{-1}  \left( \frac{\partial \bq}{\partial \bx} \right).
\end{equation}
Combining \eqref{eq:dldx} and \eqref{eq:dtdx} yields an expression for the gradient 
\begin{equation} \label{eq:adjointstatemethodgradient}
    \nabla_{\bx} \ell = \frac{\partial \ell}{\partial \bx}  - \left( \frac{\partial \ell}{\partial \bt} \right) \left( \frac{\partial \bq}{\partial \bt} \right)^{-1}  \left( \frac{\partial \bq}{\partial \bx} \right).
\end{equation}
Rewriting in terms of CGFs and their derivatives, we have
\begin{align}
    \frac{\partial \ell}{\partial \bx} \ \ = \ \ &\oner^T \left(   \frac{\partial}{\partial \bx}K_{\bG \bx + \bfeta}(\bt) - \frac 1 2 \left\{ \text{diag}\left( K''_{\bG \bx + \bfeta}(\bt) \right)  \right\}^{-1} \frac{\partial}{\partial \bx}K''_{\bG \bx + \bfeta}(\bt)    \right)_{_{}}    \label{eq:dldx2}, \\
    \frac{\partial \ell}{\partial \bt} \ \ = \ \  & \left( K'_{\bG \bx + \bfeta}(\bt) - \frac 1 2 \left( K'''_{\bG \bx + \bfeta}(\bt) \oslash K''_{\bG \bx + \bfeta}(\bt) \right) - \by  \right)^T _{_{}} \label{eq:dldt}, \\
    \frac{\partial \bq}{\partial \bt_{_{}}}  \ \ = \ \  & \text{diag}\left( K''_{\bG \bx + \bfeta}(\bt) \right)_{_{}} \label{eq:dqdt}, \\ 
    \frac{\partial \bq}{\partial \bx} \ \ = \ \  & \frac{\partial}{\partial \bx}  K'_{\bG \bx + \bfeta}(\bt), \label{eq:dqdx}
\end{align}
where $\bt$ must be found for a given $\bx$ by solving $\bq(\bx, \bt) = \bzero$. With enough patience, these derivatives can generally be calculated by hand. We present gradients for the example problems in Section \ref{sec:example} and the appendices.

%%%%%%%%%%%%%%%%%%%%%%%%%%%%%%%%%%%%%%%%%%%%%%%%%%%%%%%%%%%%%%%%%%%%%
%%%%%%%%%%%%%%%%%%%%%%%% ALGORITHMS %%%%%%%%%%%%%%%%%%%%%%%%
\section{Algorithms} \label{sec:algorithms}
So far, we proposed a method for constructing an approximate log-LF using noise from general distributions and discussed methods for calculating its gradient. Success rests on our ability to efficiently and accurately optimize $\ell$ as cast generically in \eqref{eq:optimization_problem}. There are several computational challenges for maximizing the approximate log-LF. 

First, most optimization algorithms require access to the objective's gradient. Although we present a method for obtaining a gradient when $\bt$ must be determined numerically, it can be a challenging task. For those who prefer not to toil endlessly over a cruel calculus exercise, automatic differentiation software can be employed. {ADiMat} is a popular package for {MATLAB} \cite{Bischof2002CST}. It supports reverse-mode differentiation (known as back-propogation in the context of training neural nets) which is desirable for scalar objectives of many variables. A list of packages for different languages can be found at \url{http://www.autodiff.org}.

The second challenge follows from the non-concavity of $\ell$ in $\bx$, since most algorithms can only guarantee convergence to stationary points or, at best, local maximizers, and globalization strategies are either heuristics or computationally infeasible. This is a challenge for most MLE problems, not just our formulation. Experimental evidence using multiple initializations suggests that the non-concavity effect is quite mild, especially when initializing with a reasonable guess. We found that using the OLS estimator works well as an initial guess, as it is cheap to compute and more robust than the TLS estimator.

As the problem is unconstrained with a smooth objective function, we use the well-known quasi-Newton method L-BFGS \cite{LBFGS} via the MATLAB package \texttt{minFunc}~\cite{minFunc} because L-BFGS converges more quickly than gradient descent yet doesn't require the second-order derivative information nor matrix inversion of Newton's method. Since \texttt{minFunc} is a minimization routine, we minimize $-\ell(\bx)$ in practice rather than maximimizing $\ell(\bx)$.  L-BFGS performs well on moderate or even large problems. Very large problems could be approached via stochastic gradient methods, using a subset of data to estimate a gradient, but such implementations are beyond the scope of this paper.

%%%%%%%%%%%%%%%%%%%%%%%%%%%%%%%%%%%%%%%%%%%%%%%%%%%%%%%%%%%%%%%%%%%%%
%%%%%%%%%%%%%%%%%%%%%%%% EXAMPLES %%%%%%%%%%%%%%%%%%%%%%%%

%%%%%%%%%%%%%%%%%%%%%%%% ROUNDING EXAMPLES %%%%%%%%%%%%%%%%%%%%%%%%
\section{Examples} \label{sec:example}
The following examples consider the generative model, $\by = \bG \bx + \bfeta$. Although we are free to use additive noise from other distributions depending on the problem, we elected to use Gaussian noise for clarity of exposition. In particular, $\bfeta \sim \mathcal{N}(\bzero, \sigma^2 \bI)$. The corresponding CGFs and their derivatives are
\begin{equation}
    K_{\eta_i}(t_i) = \sigma^2 t_i^2/2, \qquad K'_{\eta_i}(t_i) = \sigma^2 t_i, \qquad  \ K''_{\eta_i}(t_i) = \sigma^2.
\end{equation}

The log-LF for each example is derived and provided below. All the corresponding derivatives used for gradient calculations can be found in the appendices. We remind the reader that although analytic expressions for the gradient are possible, automatic differentiation eliminates the need for messy calculations. 

\subsection{Rounding error} \label{subsec:roundingerror}
Suppose that we observe $\by$ and a rounded version of $\bG$, denoted by $\bH$. For concreteness, assume all elements of $\bG$ are rounded to the ones spot, e.g., $1.4 \rightarrow 1$. In this case, we can model $\bG$ as a uniform random matrix of mean $\bH$. Parameter values specifying the support of the uniform random variables can be inferred from the data. If elements are rounded to the ones spot, then the uncertainty parameter will be $\delta = 0.5$; if rounded  to the tens, it will be $\delta = 5$, etc. Gaussian additive noise of known variance is assumed for $\bfeta$.

To highlight the difficulties of forming an exact likelihood for this problem, we note that $\by$ is the joint density of linear combinations of uniform and Gaussian random variables. The Irwin-Hall distribution covers the case of i.i.d. uniform random variables on $(0,1)$ which has a piece-wise polynomial PDF. Weighted sums of uniform random variables were studied in \cite{kamgar1995distribution} where the authors showed that the density can be written as a sum of polynomials through extensive use of the Heaviside function. Unfortunately, the number of terms in the expansion grows exponentially; a LF of 30 variables contains over a billion terms. The PDF for the sum of uniforms must still be convolved with a Gaussian density to obtain the true density for a single component of $\by$. With $\by \in \R^m$, a single likelihood evaluation requires summing a billion terms $m$ times! Needless to say, the problem is intractable using an exact likelihood.

The approximate likelihood presents a simple and appealing alternative. As shown in \eqref{eq:elementwiseloglikelihood}, we can focus on element-wise, univariate, random variables to form our approximate log-LF. The generating functions used to construct the LF can be found in most statistical texts. In particular, the CGF for $G_{ij} \sim \text{Uniform}(H_{ij} - \delta, H_{ij} + \delta)$ can be simplified to
\begin{equation}
    K_{G_{ij}}(t_i x_j) = t_i x_j H_{ij} + \ln\left(\frac{\sinh(\delta t_i x_j)}{\delta t_i x_j}\right), 
\end{equation}
We also require the corresponding derivatives given by 
\begin{align}
    K'_{G_{ij}}(t_i x_j) &= H_{ij} x_j - \frac{1}{t_i} + \delta x_j \coth(\delta t_i x_j), \nonumber \\
    K''_{G_{ij}}(t_i x_j) &= \frac{1}{t_i^2} - \delta^2 x_j^2 \csch^2(\delta t_i x_j).
\end{align}
Putting it together gives the log-LF
\begin{align}
    \ell(\bx) = \sum_{i=1}^m \left\{ \rule{0cm}{.75cm} \right. \frac{\sigma^2 t_i^2}{2} +  \sum_{j=1}^n  & \Bigg[ H_{ij} t_i  x_j \ + \ \ln \Bigg(\frac{\sinh(\delta t_i x_j)}{\delta t_i x_j} \Bigg) \Bigg] - \ t_i y_i   \\
    -  & \ \frac{1}{2} \ln  \left[ \sigma^2  + \sum_{j=1}^n \left( \frac{1}{t_i^2} - \delta x_j^2 \csch^2(\delta t_i x_j) \right)   \right] \left. \rule{0cm}{.75cm} \right\}  \nonumber
\end{align}
and can be written in matrix/vector form as 
\begin{align}
    \ell(\bx) = \bt^T \left( \frac{\sigma^2}{2} \, \bt +  \bH \, \bx - \by \right) & + \oner^T \, 
    \ln \left[ \sinh  \left(\dtx \right) \oslash \left(\dtx \right)    \right] \, \oner   \\
    &- \frac{1}{2}\oner^T \, \ln \left[\sigma^2 \, \oner + n \oslash \bt^2 - \delta^2 \, \csch^2\left(\dtx \right) \bx^2 \right],  \nonumber
\end{align}
where $\bt$ solves the equation $K'_{\bG \bx + \bfeta}(\bt) = \by$. 

The above example is principled and can be used in instances of rounding. The observed design matrix is used directly since it (along with $\delta$) completely specifies the distribution from which our ``true'' design matrix was drawn. The distribution of $\bG$ and all its parameters can be inferred by observing $\bH$. 

\subsection{Floating point uncertainty}
For the case of floating point uncertainty, all appearances of $\delta$ should be replaced by $D_{ij}$ or $\bD$ depending on whether it appears in an element-wise or matrix/vector equation, respectively. Specifically, by letting $\bD$ be a non-negative matrix specifying the uncertainty based on the number of significant figures included. For example, if $H_{ij} = \num{1.7e4}$ then $D_{ij} = \num{0.05e4}$. Setting $\bM = \bD \odot (\bt \bx^T)$, we have 
\begin{align}
    \ell(\bx) = \bt^T \left( \frac{\sigma^2}{2} \, \bt +  \bH \, \bx - \by \right) & + \oner^T \, 
    \ln \left[ \sinh  \left( \bM \right) \oslash  \bM     \right] \, \oner   \\
    &- \frac{1}{2}\oner^T \, \ln \left[\sigma^2 \, \oner + n \oslash \bt^2 - \left( \bD^2 \odot \csch^2\left(\bM\right) \right) \bx^2 \right].  \nonumber
\end{align}
The derivatives provided in the appendix are for the floating-point uncertainty which reduces to fixed-point or rounding error by letting $\bD = \delta \oner^{\,}_m \oner_n^T$.

%%%%%%%%%%%%%%%%%%%%%%%% EXPONENTIAL EXAMPLE %%%%%%%%%%%%%%%%%%%%%%%%
\subsection{Exponential clipping}
Suppose that the the elements of $\bG$ are drawn from a double exponential distribution with rate $\lambda$ and that entries with magnitudes greater than some threshold, $\gamma$ say, are clipped. The clipped matrix is given by
\[
    \bH = \sign(\bG) \cdot \min \{ |\bG|, \gamma \},
\]
with $\min\{\cdot , \gamma \}$ and the absolute values operating element-wise. In this case, uncertainty is only realized for clipped components. Let $\bS = \sign(\bH)$ and $\bA$ be defined by
\[
    A_{ij} = \begin{cases} 1, \quad |H_{ij}| =\gamma, \\
    0, \quad \text{else.}
    \end{cases}
\]
By the memorylessness property of exponentials, the uncertainty of clipped entries will also be $\pm$Exponential$(\lambda)$. Using Gaussian additive noise, the remaining CGFs and corresponding derivatives are given by 
\begin{align}
    K_{G_{ij}}(t_i x_j) &= t_i H_{ij} x_j - A_{ij} \, \ln \left( 1 - \frac{S_{ij} t_i x_j}{\lambda}\right), \nonumber \\  
    K'_{G_{ij}}(t_i x_j) &=  H_{ij} x_j + \frac{A_{ij} S_{ij} x_j}{\lambda - S_{ij} t_i x_j},  \\
    K''_{G_{ij}}(t_i x_j) &= \frac{ A_{ij}  x_j^2}{(\lambda - S_{ij} t_i x_j)^2} . \nonumber
\end{align}
Using the CGFs above, the approximate log-LF is given by
\begin{align}
    \ell(\bx) = \sum_{i=1}^m \left\{ \rule{0cm}{.75cm} \right. \frac{\sigma^2 t_i^2}{2} + & \ \sum_{j=1}^n \Bigg[ t_i H_{ij} x_j - A_{ij} \, \ln \left( 1 - \frac{S_{ij} t_i x_j}{\lambda}\right) \Bigg]   \\
    - & \ t_i y_i -  \frac{1}{2} \ln \left[ \sigma^2 - \sum_{j=1}^n \frac{A_{ij} \, x_j^2}{(\lambda - S_{ij} t_i x_j)^2}    \right] \left. \rule{0cm}{.75cm} \right\}. \nonumber
\end{align}
Letting $\bLam = \lambda \oner_m^{\,} \oner_n^T \in \R^{\mbyn}$ and $\bC = \bS \odot (\bt \bx^T)$ the matrix/vector form is
\begin{align}
    \ell(\bx) = \bt^T \left( \frac{\sigma^2 \bt}{2} + \bH \bx - \by \right) +& \oner^T \left(  \bA \odot  \ln \left[ \bLam \oslash \left(\bLam - \bC \right) \right] \right) \oner \\ 
    -&\frac{1}{2} \oner^T \ln \left\{    
    \sigma^2 \oner + \left[  \bA \oslash (\bLam -  \bC )^2 \right] \bx^2
    \right\} \nonumber.
\end{align}
Derivatives for construction of the gradient can be found in the appendix. 

It can be observed that the approximate log-LF above and its derivatives have singularities depending on component values of $\bx$. Since root finding algorithms generally require continuity, complications arise when solving for the $\bt$ that verifies $K'_{\bG \bx + \bfeta} (\bt) = \by$. Consequently, a modified root-finding algorithm must be employed that brackets a continuous interval about zero.

%%%%%%%%%%%%%%%%%%%%%%%% GAUSSIAN EXAMPLE %%%%%%%%%%%%%%%%%%%%%%%%
\subsection{Gaussian uncertainty}
For our final example, both the design matrix and measurement vector are subject to Gaussian noise.
In this case, there exists an exact closed form likelihood that can be derived directly from the joint density and we show that our method recovers the exact log-LF. Suppose that $\bG$ is Gaussian with mean $\bH$ and element-wise variance $\rho^2$. Similarly, we let $\bfeta \sim \mathcal{N}(\bzero, \sigma^2\bI)$. Our CGFs are
\begin{equation} \label{eq:gaussiancgfs}
    K_{\eta_i}(t_i) = \frac{1}{2}\sigma^2 t_i^2 \qquad \tand \qquad K_{G_{ij}}(t_i x_j) = \bh_i^T \bx  \, t_i + \frac{1}{2}\rho^2 t_i^2 x_j^2.
\end{equation}
Recalling that $t_i$ is the solution to $K_{\bg_i^T \bx + \eta_i}'(t_i) = y_i$, we can solve for $t_i$ giving 
\begin{equation} \label{eq:gaussianroots}
    t_i = \frac{y_i - \bh_i^T \bx}{\sigma^2 + \rho^2 \|\bx \|^2}.
\end{equation}
Plugging \eqref{eq:gaussiancgfs} and \eqref{eq:gaussianroots} into \eqref{eq:elementwiseloglikelihood} gives our ``approximate'' log-LF,
\begin{equation}
    \ell(\bx) = -\frac{1}{2} \left[  \frac{\|\by - \bH \bx \|^2}{\sigma^2 + \rho^2 \|\bx \|^2} + m \, \ln \left(\sigma^2 + \rho^2 \|\bx \|^2\right) \right].
\end{equation}
By solving for $\bt$ as a function of $\bx$, the gradient is given directly by 
\begin{equation}
    \nabla_{\bx} \ell(\bx) = \frac{1}{\sigma^2 + \rho^2 \|\bx\|^2} \left[
        \rho^2 \left(      
            \frac{\|\bH \bx - \by\|^2}{\sigma^2 + \rho^2 \|\bx\|^2} - m
        \right) \bx - \bH^T ( \bH \bx - \by ) 
    \right].
\end{equation}

We note that $\ell(\bx)$ is the \textit{exact} log-likelihood, up to a constant, as the one derived in \cite{wiesel2008linear}. This follows from the fact that a Gaussian is uniquely determined by its first and second cumulants, i.e., mean and variance. The truncated Taylor series used for the saddle point approximation is precise by virtue of all higher-derivatives being zero.

%%%%%%%%%%%%%%%%%%%%%%%%%%%%%%%%%%%%%%%%%%%%%%%%%%%%%%%%%%%%%%%%%%%%%
%%%%%%%%%%%%%%%%%%%%%%%% EXPERIMENTS %%%%%%%%%%%%%%%%%%%%%%%%
\section{Numerical experiments} \label{sec:experiments}
\begin{figure}
    \centering
    \includegraphics[width=.6 \textwidth]{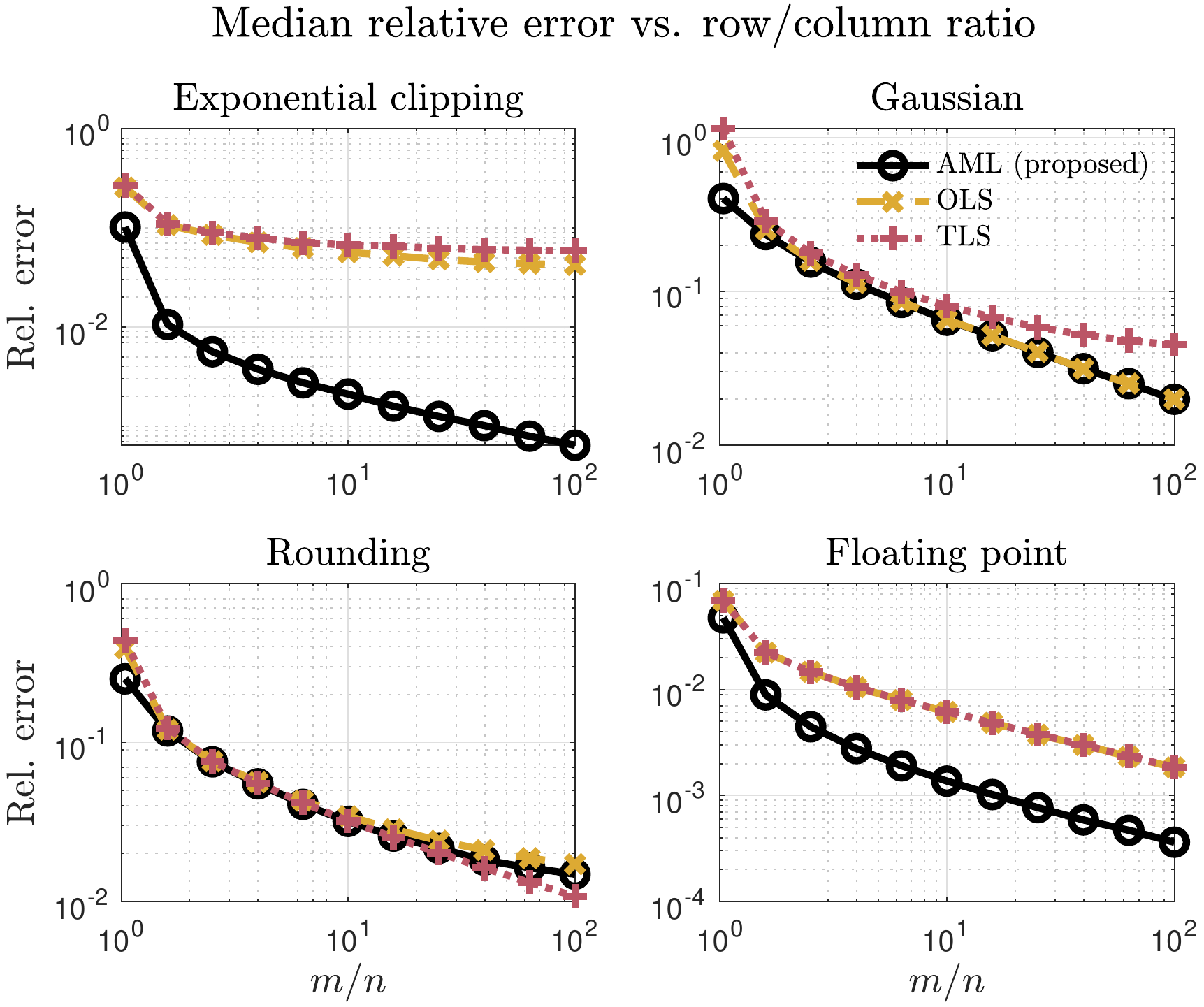} 
    \caption{Median relative error $\left(  \frac{\|\xest - \xtrue\|}{\|\xtrue\|} \right)$ over 1,000 simulations. Number of columns fixed at $n=20$ while row count varied from $m=21$ to $m=2,000$ showing estimator performance with additional data.} 
    \label{fig:varyrows}
\end{figure}

To validate the approximate likelihood estimator, we implemented our proposed method along with OLS and TLS on the model examples discussed in Section \ref{sec:example}. Although it is possible to use a weighting matrix for TLS to account for non-uniform variance between the operator and measurement vector, we observed little benefit and opted to solve the unweighted version. In each case we used the generative model $\by = \bG \xtrue + \bfeta$ with $\bfeta \sim \mathcal{N}\left(\bzero, (0.1)^2 \bI \right)$. The variance on the additive Gaussian noise 
was chosen as to not overwhelm uncertainty in $\bG$. The ``true'' solution vectors, denoted by $\xtrue$, were drawn from a heavy-tailed Cauchy distribution; this distribution makes the use of prior information on the solution difficult. Although $\xtrue$ is randomly drawn, it is fixed for each simulation and does not introduce uncertainty into the problem. The vector $\by$ is observed.

For each example, the matrices in question were created as follows:
\begin{itemize}
    \item \textbf{Rounding}: The elements of $\bG$ were drawn from a continuous uniform distribution supported on $(0,10)$. The observed matrix $\bH$ was constructed by rounding $\bG$ to the ones spot, i.e., $\bH = \text{Round}(\bG)$ with uncertainty parameter $\delta = 0.5$. 
    
    \item \textbf{Floating point}: The elements of $\bG$ are the product of a standard normal Gaussian and $10^k$ where $k$ is uniformly chosen from $\{0,1,2,3\}$. The observed matrix $\bH$ is given by keeping two significant figures. The uncertainty for a particular element is scaled correspondingly, i.e., if $H_{ij} = \num{-3.1e2}$ then $D_{ij} = \num{0.05e2}$.
    
    \item \textbf{Exponential clipping}: The elements of $\bG$ were drawn from a double exponential or Laplace distribution with rate $\lambda = 2$. Fixing the clipping threshold at $\gamma =2$, matrix $\bH = \sign(\bG) \max\{ | \bG|, \gamma \}$ is observed. We note that for $\lambda = \gamma = 2$, approximately $2\%$ of matrix elements are clipped. 
    
    \item \textbf{Gaussian}: A mean observed matrix $\bH$ was drawn randomly from a Gaussian of variance 100 (this merely fixes a mean and could be chosen from an arbitrary distribution). The elements of matrix $\bG$ were then drawn from a Gaussian of mean $\bH$ and variance $\rho^2 = 4$. The Gaussian approximate MLE corresponds to the exact MLE. 
\end{itemize}
For each listed problem, it is assumed that $\bH, \ \by$, and the relevant distributional parameters are known. We conducted three simulations for each example instance:
\begin{enumerate}
    \item Fixed the number of columns at $n=20$ then allowed the number of rows to increase from $m=21$ to $m=2000$ reflecting how the estimator responds to more data.
    \item Fixed the number of rows at $m=100$ then allowed the number of columns to increase from $n=1$ to $n=99$ evaluating how the estimator responds to an increasing number of unknowns. 
    \item For $m=55$ and $n=50$, we investigated the distribution of errors the approximate MLE error compares to that of OLS and TLS. 
\end{enumerate}
The approximate MLE, OLS, and TLS estimators are denoted by $\xmle$, $\xols$, and $\xtls$, respectively. A generic estimator is denoted by $\xest$.

\begin{figure}
    \centering
    \includegraphics[width=.60\textwidth]{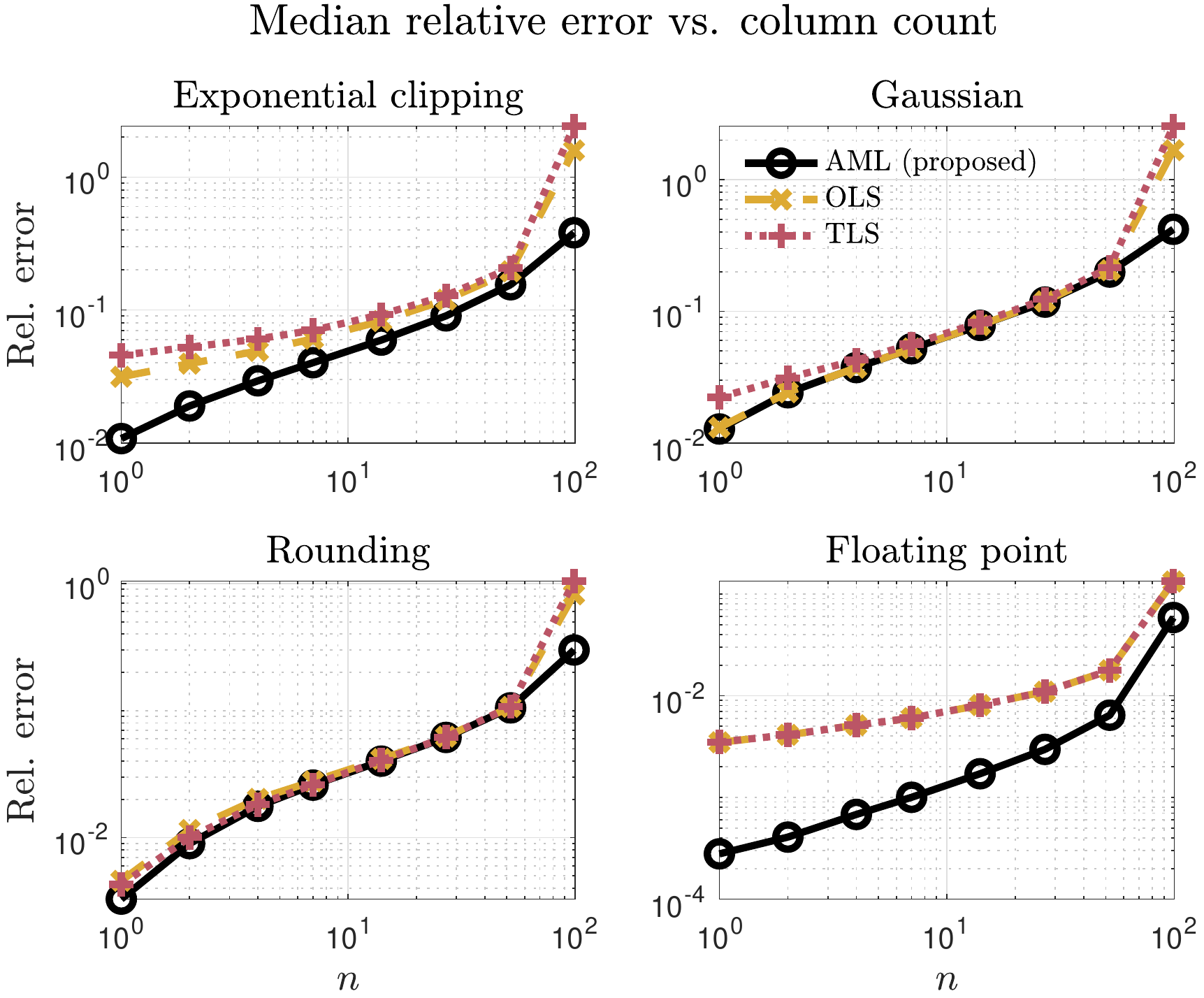}
    \caption{Median relative error $\left(  \frac{\|\xest - \xtrue\|}{\|\xtrue\|} \right)$ over 1,000 simulations. Number of rows fixed at $m=100$ while row count varied from $n=1$ to $n=99$ showing estimator performance for an increasing number of unknowns.}
    \label{fig:varycolumns}
\end{figure}

The first two simulations summarized in Figs. \ref{fig:varyrows} and \ref{fig:varycolumns} show that AML outperforms the other methods when the system is only slightly over determined. The method also excels in the highly over-determined regime for exponential clipping and floating point uncertainty. The relatively small extent of operator uncertainty as is typically encountered in the case of rounding error might explain why the proposed method fails to distinguish itself there. It is worth noting similar behavior for Gaussian uncertainty which corresponds to the \emph{exact} MLE. As a result, we believe the modest performance stems from the limited scale of uncertainty rather than the method itself.

\begin{figure}
    \centering
    \includegraphics[width=.48\textwidth]{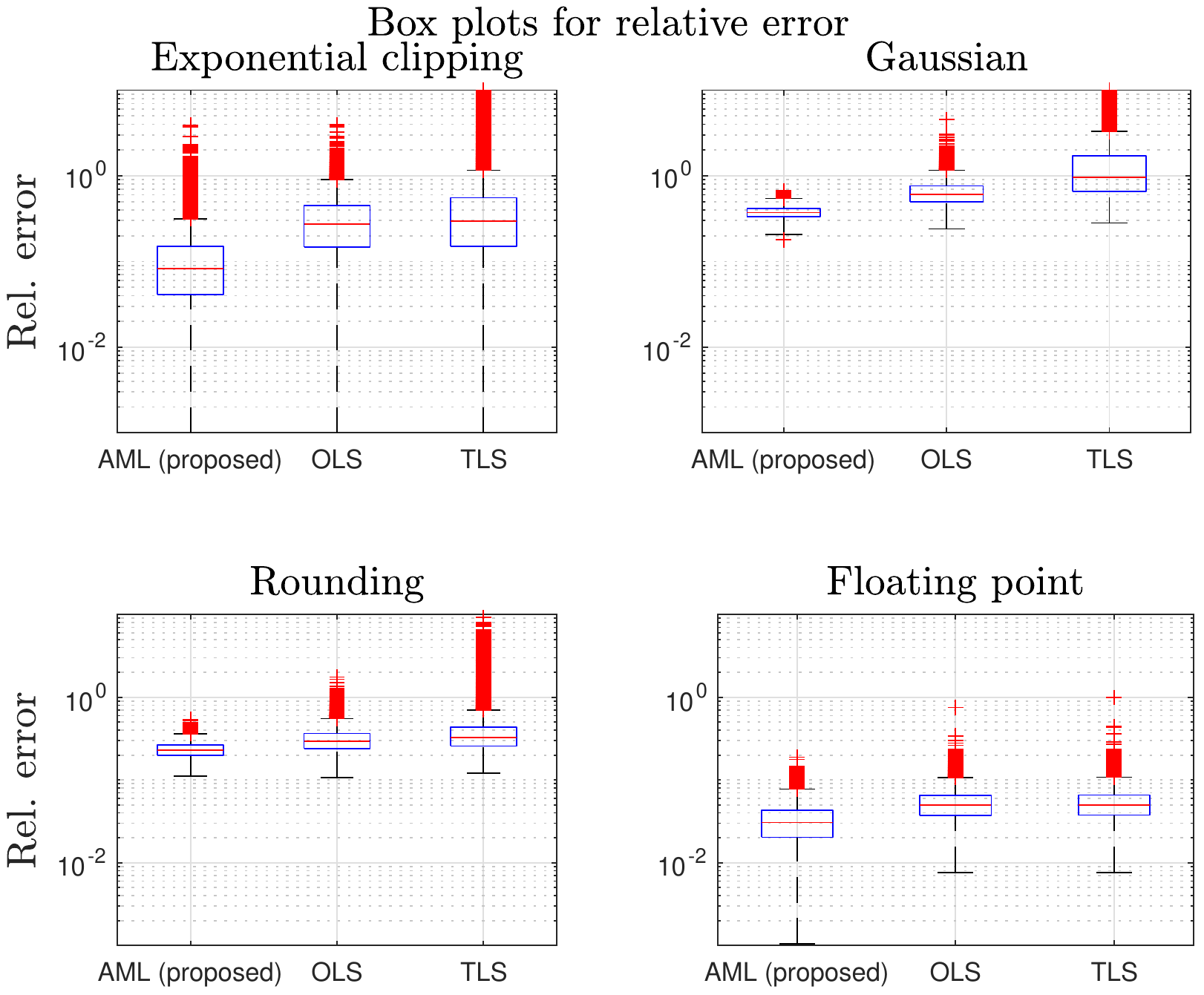}
    \includegraphics[width=.48\textwidth]{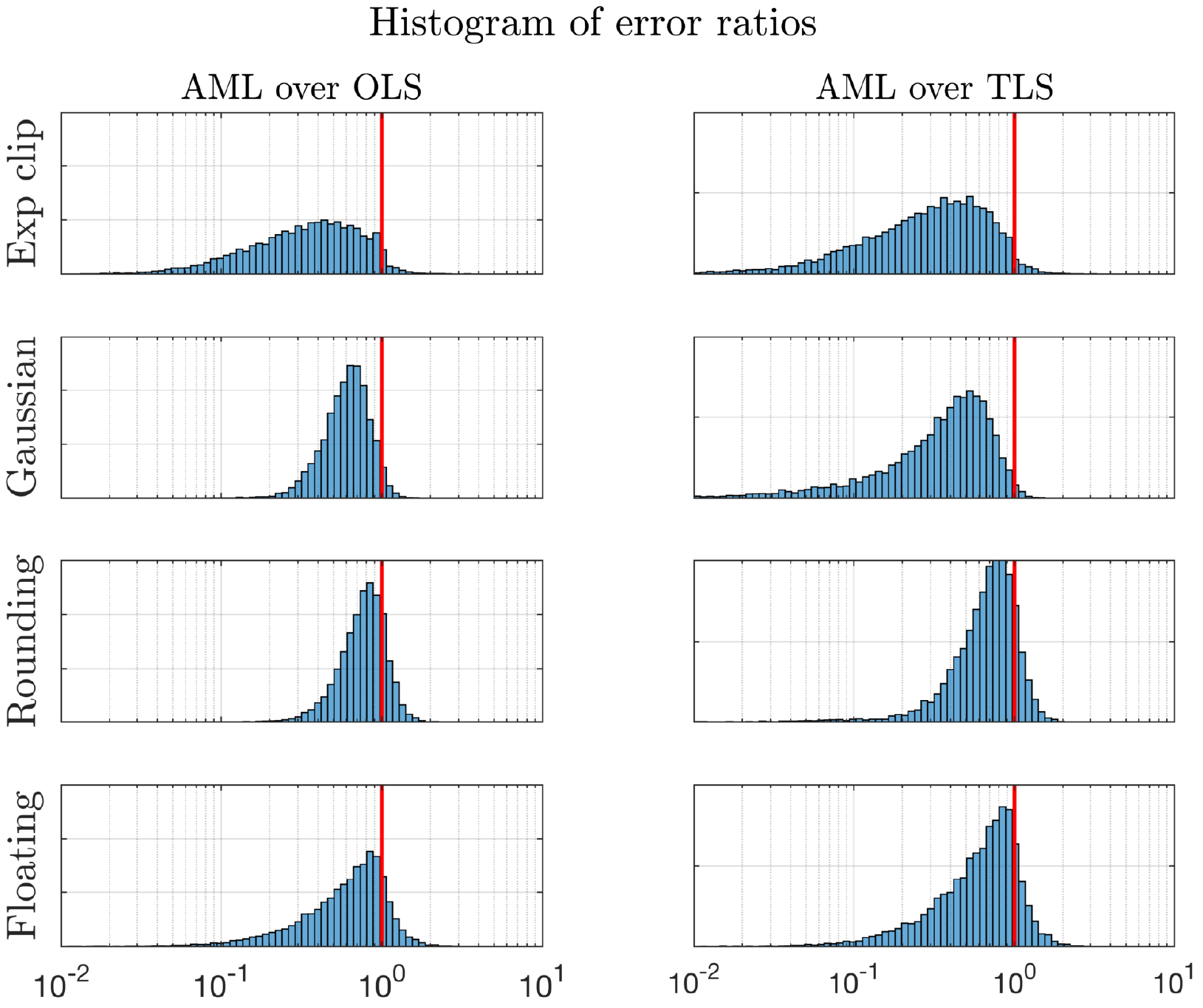}
    \caption{Error metrics for simulations $\bG \in \R^{55 \times 50}$ over $10,000$ simulations. Left: box-plots of relative error for different methods, i.e., $\frac{\|\xtrue - \xest \|}{ \| \xtrue\|}$. Box indicates middle $50$th quantile with interior line at the median. Bars denote extreme values with outliers indicated by ``+''. Right: histogram of error ratio $\frac{\| \xmle - \xtrue \|}{ \| \xols - \xtrue \|}$ and $\frac{\| \xmle - \xtrue \|}{ \| \xtls - \xtrue \|}$. Values less than 1 (vertical line) indicate AML outperformed competing method for identical data.}
    \label{fig:boxplothist}
\end{figure}

The large benefit of AML for limited data is further illustrated in simulation 3 as shown in Fig. \ref{fig:boxplothist}. The figure gives box plots for each model and the error ratio or relative improvement of AML over OLS and TLS, i.e., $\frac{\|\xmle - \xtrue\|}{\|\xols  - \xtrue\|}$ and $\frac{\|\xmle - \xtrue\|}{\|\xtls  - \xtrue\|}$, respectively. For the histogram, all values less than one (red line) are improvements over competing estimator for the same problem instance. Although AML does not always outperform the alternative methods, it usually does with pronounced gains. The code use for simulations can be found at \url{https://github.com/rclancyc/approximate_mle}.

\section{Conclusion}
In this work, we cast a regression problem in the maximum likelihood estimation framework. We discussed difficulties associated with forming a true and exact likelihood function for models with noise in the design matrix, then presented a method for generating an approximate log-LF based on the saddle point approximation. The proposed log-LF is easily constructed using component-wise moment generating functions and a simple root-finding algorithm. General gradient calculations were presented to help efficiently solve the resultant optimization problem. Quasi-Newton methods performed well in our numerical experiments. 

Attention was paid primarily to the case of exponential clipping, Gaussian noise, rounding, and floating point error which motivated the authors initially and provide a principled example where the distributions involved were easily inferred. Despite our focus on several examples, the method remains widely useful, particularly when noise in the design matrix comes from unwieldy distributions with difficult to handle linear combinations. We envision the proposed method finding use anytime the design matrix is subject to ambiguity. In subsequent work, we plan to establish probabilistic guarantees on distance of approximate MLEs to true MLEs and further investigate behavior for different noise distributions.

\newpage

\bibliographystyle{ieeetr}
\bibliography{references}

\newpage

\appendix
\section{Gradient factors}
The gradient for our approximate likelihood function is outlined in Equations \ref{eq:dldx2}-\ref{eq:dqdx} when $\bt$ cannot be solved for explicity. To calculate $\nabla_{\bx} \ell(\bx)$, we need expressions for the joint CGF and a number of its derivatives. In particular, we require $K^{(i)}_{\bG \bx + \bfeta}(\bt)$ for $i = 0,1,2,3$ and $\frac{\partial}{\partial \bx} K^{(j)}_{\bG \bx + \bfeta}(\bt)$ for $j = 0,1,2$. We present the necessary derivatives for gradient determination for the examples listed in Section \ref{sec:example}. Recall that $\bt$ is the solution to the equation $K'_{\bG \bx + \bfeta}(\bt) = \by$ for a given $\bx$. 

\subsection{Floating point/rounding error}
In what follows, let $\bD$ be the floating point error matrix and $\bM = \bD \odot (\bt \bx^T)$. For fixed point or rounding error with uncertainty parameter $\delta$, the matrix reduces to $\bD = \delta \oner^{\,}_m \oner^T_n$ such that $\bM = \delta \bt \bx^T$
\begin{align}
    K_{\bG \bx + \bfeta}(\bt) \ \ = \ \ & \frac{\sigma^2 \bt^2}{2} + \bt \odot \bH \bx + \ln \left( \sinh\left( \bM \right) \oslash  \bM  \right) \oner \\
    K'_{\bG \bx + \bfeta}(\bt) \ \ = \ \ & \frac{}{} \sigma^2 \bt + \bH \bx + \left[\bD \odot \coth\left(\bM  \right) \right] \bx - n (\oner \oslash \bt) \\
    K''_{\bG \bx + \bfeta}(\bt) \ \ = \ \ & \frac{}{} \sigma^2 \oner - \left[\bD^2 \odot \csch^2 \left( \bM \right)\right] \bx^2 + n (\oner \oslash \bt^2) \\
    K'''_{\bG \bx + \bfeta}(\bt) \ \ = \ \ & \frac{}{} 2 \left[ \bD^3 \odot \coth\left( \bM \right) \odot \csch^2 \left( \bM \right) \right] \bx^3 - 2n(\oner \oslash \bt^3).
\end{align}
Now differentiating the CGF and derivatives with respect $\bx$ we have
\begin{align} \label{eq:appendix_equations}
    \frac{\partial}{\partial \bx} K_{\bG \bx + \bfeta}(\bt) \ \ = \ \ & \bt \oner^T \odot \left[ \bH + \bD \odot \coth\left( \bM \right)   \right] - \oner \left(\oner \oslash \bx\right)^T \\
    \frac{\partial}{\partial \bx} K'_{\bG \bx + \bfeta}(\bt) \ \ = \ \ & \bH + \bD \odot \coth \left( \bM \right) - \bD \odot \bM \odot \csch^2 \left( \bM \right)  \\
    \frac{\partial}{\partial \bx} K''_{\bG \bx + \bfeta}(\bt) \ \ = \ \ & 2 \bD^2 \odot \csch^2 \left( \bM \right) \odot \left[ \left( \bD \odot \bt \left( \bx^2 \right)^T \right) \odot \coth\left( \bM \right)  - \oner 
     \bx^T \right].
\end{align}

\subsection{Exponential clipping} 
 In the case of exponential clipping, $\lambda$ is the rate of the exponential for which elements of $\bG$ are drawn. Letting $\bLam = \lambda \oner^{\,}_m \oner_n^T$ and $\bC = \bS \odot (\bt \bx^T)$ derivative are given as
\begin{align}
    K_{\bG \bx + \bfeta}(\bt) \ \ = \ \ & \frac{\sigma^2 \bt^2}{2} + \bt \odot \bH \bx + 
    \left( \bA \odot    
    \ln \left[
         \bLam \oslash \left(
            \bLam - \bC
         \right)
    \right]
    \right) \oner,
    \\
    K'_{\bG \bx + \bfeta}(\bt) \ \ = \ \ & \frac{}{} \sigma^2 \bt + \bH \bx +
    \left[
    \bA \odot \bS \oslash (\bLam - \bC)
    \right] \bx,
    \\
    K''_{\bG \bx + \bfeta}(\bt) \ \ = \ \ & \frac{}{} \sigma^2 \oner +
    \left[
    \bA \oslash (\bLam - \bC)^2
    \right] \bx^2,
    \\
    K'''_{\bG \bx + \bfeta}(\bt) \ \ = \ \ & \frac{}{} 2 \,
    \left[
    \bA \odot \bS \oslash (\bLam - \bC)^3
    \right] \bx^3.
\end{align}
Now differentiating the CGF and derivatives with respect $\bx$ we have
\begin{align}
    \frac{\partial}{\partial \bx} K_{\bG \bx + \bfeta}(\bt) \ \ = \ \ & 
    \left(\bt \oner^T \right) \odot \left[ \bH + \bA \odot \bS \oslash \left( \bLam - \bC  \right) \right]
    \\
    \frac{\partial}{\partial \bx} K'_{\bG \bx + \bfeta}(\bt) \ \ = \ \ & 
    \bH + \bA \odot \left[ \bS \oslash (\bLam - \bC) +  \left(\bt \bx^T \right) \oslash (\bLam - \bC)^2 \right]
    \\
    \frac{\partial}{\partial \bx} K''_{\bG \bx + \bfeta}(\bt) \ \ = \ \ & 
    2 \bA \odot \left[ \left(\oner \bx^T \right) \oslash (\bLam - \bC)^2 + \bS \odot \left(\bt \left( \bx^2 \right)^T \right) \oslash (\bLam - \bC)^3 \right].
\end{align}

\end{document}